\documentclass[reprint,amsmath,amssymb,aps,prd,nofootinbib]{revtex4-2}

\usepackage[margin=0.95 in]{geometry} 
\usepackage{scalerel}

\newcommand{\curlyD}{\mathcal{D}}

\newcommand{\curlyH}{\mathcal{H}}
\newcommand{\curlyS}{\mathcal{S}}
\newcommand{\II}{\mathcal{I}}

\newcommand{\pd}{\partial}
\begin{document}

\title{Grand-Canonical Symmetric Orbifold Theories}
\author{Lior Benizri}
\affiliation{Laboratoire de Physique de l'\'Ecole Normale Sup\'erieure$,$ CNRS$,$ ENS$,$ Universit\'e PSL$,$  Sorbonne Universit\'e$,$ 
Universit\'e  Paris Cit\'e F-75005 Paris$,$ France
}

\begin{abstract} 
In this letter, we study grand-canonical symmetric orbifolds of conformal field theories. We propose to define them as the direct sum of symmetric orbifolds of all degrees. The natural basis of operators is one that mixes all sectors. We describe this basis in terms of partial permutations, and explain how to define and calculate the operator product expansion in that framework. Our construction provides a conformal field theory interpretation of the central charge operator $\II$ in $AdS_3$ string theory.
\end{abstract}

\maketitle

\section{Introduction}
String theory on an $AdS_3\times S^3 \times T^4$ background with one unit of NS-NS flux is dual to a grand-canonical ensemble of symmetric orbifold conformal field theories \cite{Eberhardt:2018ouy,Eberhardt:2019ywk,
Kim:2015gak,Eberhardt:2020bgq,Eberhardt:2021jvj}. So far, these ensembles have been defined solely in terms of their correlation functions. Up to a normalization factor, these read
\begin{equation}
    \langle \cdot \rangle_p := \sum_d p^d \langle \cdot \rangle^{\rm un}_{d}\, ,
\end{equation}
where $\langle \cdot \rangle^{\rm un}_{d}$ denotes the unnormalized correlation function in the symmetric orbifold of degree $d$. The purpose of this letter is to clarify the structure of these ensembles. We will define them as the direct sum of symmetric orbifolds of all degrees, thus enlarging their Hilbert space beyond that of a single symmetric orbifold in the large $N$ limit. This immediately implies that they are consistent, conformally invariant quantum field theories. The basis of operators of physical interest is one that mixes all sectors, and we will focus on spelling out its properties. In particular, we will explain how to take operator product expansions directly in this basis. We will be especially interested in the combinatorial properties of the operator that maps to the central charge operator $\II$ of the dual string theory. 

\subsection{Symmetric orbifolds}
Symmetric orbifold theories are defined by tensoring $d$ copies of a two-dimensional seed conformal field theory and gauging the group $S_d$ that permutes the copies. The identification of configurations which differ only by a permutation of the copies introduces new sectors that obey twisted boundary conditions. In order to deal efficiently with these new sectors, we introduce twist fields $\sigma_g$ that generate the twisted vacua and thus modify the boundary conditions obeyed by the fields of the theory. We write
\begin{equation}
    X_I(e^{2\pi i} z) \sigma_{g}(0) = X_{g(I)}(z) \sigma_{g}(0)\, ,
\end{equation}
where $I$ labels the different copies of the field $X$. To obtain gauge-invariant twists, we sum over all permutations in a given conjugacy class $[g]$: 
\begin{equation}
    \sigma_{[g]}(x):=\sum_{g\in [g]} \sigma_{g}(x)\, .
\end{equation}
The conformal dimension of a twist operator $\sigma_k$ associated to single cycles of length $k$ is given by \cite{Lunin:2000yv}:
\begin{equation}
    h_k = \frac{c_{seed}}{24} \cdot \left( k-\frac{1}{k} \right)\, ,
\end{equation}
where $c_{seed}$ denotes the central charge of the seed theory. Furthermore, the Hilbert space of the symmetric orbifold theory is given by a direct sum over all sectors
\begin{equation}
    \curlyH = \bigoplus_{[g]} \curlyH_{[g]}\, .
    \label{protoHilbert}
\end{equation}
We note that the Hilbert space of a multi-cycle twisted sector is given by 
\begin{equation}
    \curlyH_{[g]}=\bigotimes_{n} (\underbrace{\curlyH_{n}\otimes \dots \otimes \curlyH_{n}}_{k_n\, \text{times}})^{S_{k_n}}\, ,
\end{equation}
where $k_n$ denotes the number of $n$ cycles in $g$. Moreover, the stress-energy tensor $T(z)$ of the orbifold theory is the sum of the stress tensors $T_I(z)$ of the individual copies. Accordingly, the central charge is  $d\cdot c_{seed}$. Each sector $\curlyH_n$ is generated by acting on the corresponding twisted ground state $\lvert \sigma_n \rangle$ with the fractional modes $(L_{m/n})_{m\in \mathbb{Z}}$ of the stress tensor in that sector \cite{Roumpedakis:2018tdb,Dei:2019iym}.

There are various ways to compute correlators of a two-dimensional symmetric orbifold theory, see \cite{Lunin:2000yv,Pakman:2009zz,Pakman:2009ab,Dei:2019iym}. We briefly describe here the covering space approach developed in \cite{Lunin:2000yv}. Given a manifold with non-trivial monodromies specified by a set of permutations $\{g_1,\dots,g_n\}$, we construct the corresponding covering map $\Gamma$. We thereby trade $d$ fields obeying twisted boundary conditions for a unique single-valued field that lives on the covering surface. The correlators
\begin{multline}
    \langle \sigma_{g_1}(x_1)\cdots \sigma_{g_n}(x_n)\cdot O_1(y_1)\cdots O_{m}(y_m) \rangle_d :=\\
    \frac{1}{d!} \int_{g_i-twisted} \curlyD^d X e^{-S[X]} O_1(y_1)\cdots O_{m}(y_m)\, .
    \label{twistdef}
\end{multline}
can be computed by lifting to the covering space, and gauge-invariant observables are then obtained by taking appropriate sums. This procedure requires a Weyl transformation to cancel the determinant of the covering map. The Weyl transformation modifies the measure of the path integral, conjuring a Liouville action which must be computed and appropriately regularized. The details of this procedure are not necessary to our discussion, and we therefore refer the reader to the original paper \cite{Lunin:2000yv}.

\subsection{Road map}
One motivation for studying grand-canonical ensembles of symmetric orbifold theories is the apparent lack of an operator product expansion. Consider indeed the following grand-canonical correlator:
\begin{equation}
    \langle \sigma_k(z) \sigma_{k}(0) \rangle_p := \frac{1}{Z_p}\sum_{d\geq k} p^d \langle \sigma_k(z) \sigma_{k}(0) \rangle_d\, ,
\end{equation}
where $Z_p$ denotes the partition function of the grand-canonical theory. At each degree, one can take the operator product expansion of the twist operators. However, the corresponding OPE coefficients depend on the degree. For example, $C_{kk}^{\quad 0}(d)
=\frac{d!}{k(d-k)!}$. Therefore, it is unclear whether an operator product expansion can be defined in the grand-canonical theory. To prove that such an expansion would be inconsistent, let us compute the grand-canonical two-point function explicitly on the sphere:
\begin{equation}
     \langle \sigma_k(z) \sigma_{k}(0) \rangle_p = \frac{p^k Z_{seed}^k }{k} \frac{1}{z^{2h_k}} \, ,
     \label{2pt}
\end{equation}
where $Z_{seed}$ denotes the partition function of the seed theory. We read off the OPE coefficient $g_{kk}^{\ \ 0}$:
\begin{equation}
     g_{kk}^{\ \ 0}=\frac{p^k Z_{seed}^k }{k}\, .
\end{equation}
This result is incompatible with certain four-point functions of the theory. Indeed, it implies that the contribution to a four-point function $\langle \sigma_k \sigma_k \sigma_l \sigma_l \rangle$ obtained from contracting the two $\sigma_k$ together and the two $\sigma_l$ together is
\begin{equation}
    \langle \sigma_k(x) \sigma_{k}(y) \sigma_l(z) \sigma_{l}(0) \rangle_p = \frac{p^{k+l}}{k\cdot l}\cdot Z_{seed}^{k+l} + \dots \, ,
\end{equation}
whereas a careful computation yields
\begin{equation}
    \frac{1}{k\cdot l} \frac{1}{Z_p} \sum_d p^d Z_{seed}^d \frac{d!}{(d-k)! (d-l)!}\, .
\end{equation}
This indicates that the grand-canonical theory lacks a straightforward operator product expansion.

Another line of inquiry that motivates the study of grand-canonical symmetric orbifolds comes from the non-scalar central charge in the algebra of currents of $AdS_3$ string theories with pure NS-NS flux \cite{Kutasov:1999xu}.

The central charge operator $\II$ is the integrated vertex operator of winding number $w=1$, spin zero, and conformal dimension zero. It acts inside any fixed-genus correlator as a multiple of the identity \cite{Kutasov:1999xu}, with a constant of proportionality set by the representations of the Virasoro algebra featured in the correlator \cite{Kim:2015gak}. The nature of its counterpart in the conformal field theory remains to be understood, and a general characterization likely requires further insight into the dual theory. That said, in settings where the dual field theory is known, the lack of an intrinsic conformal field theory definition of the central charge operator is striking. String theory on an $AdS_3\times S^3\times T^4$ background with one unit of NS-NS flux is dual to a grand-canonical ensemble of symmetric orbifolds with seed the torus superconformal theory. Where in this theory is the central charge operator found ?

An alternative way to describe the central charge operator $\II$ is as a differential operator with respect to the chemical potential \cite{Eberhardt:2021jvj}. As such, it generates an algebra of operators absent from the Hilbert space of the symmetric orbifold. Therefore, if the central charge operator of the string theory admits a field-theoretic dual, then its description requires the Hilbert space of the grand-canonical theory to be substantially enlarged. Following the intuition gained from the study of topological symmetric orbifolds \cite{Li:2020zwo,Benizri:2024mpx,Benizri:2025xmz}, we will propose that the algebraic structure required for this enlargement is that of partial permutations \cite{IvanovKerov}. A partial permutation is a permutation of a subset of $n$ elements. It is composed of a support $\curlyS\subset \{1,\dots, n\}$ and a permutation $\rho$ of $\curlyS$. We denote it as $(\curlyS,\rho)$, or alternatively, with the Latin letter $r$. The product of two partial permutations is obtained by multiplying the permutations and concatenating their supports:
\begin{equation}
    (\curlyS_1,\rho_1)\cdot (\curlyS_2,\rho_2) = (\curlyS_1\cup \curlyS_2, \rho_1\rho_2)\, .
\end{equation}
Moreover, the symmetric group $S_n$ acts on partial permutations by conjugation. We write
\begin{equation}
    \pi\cdot (\curlyS,\rho) = (\pi \curlyS,\pi \rho \pi^{-1})\, .
\end{equation}
We define the orbit sum $A_{[r]}$ as the sum over all partial permutations in the conjugation orbit of $r$. These orbit sums are characterized by the size of their support and their cycle type. Our proposal is that in a grand-canonical theory, twist operators are indexed not by orbits of the infinite symmetric group, but rather by orbits of partial permutations. An immediate consequence of this proposal is that there are many central operators. In section \ref{Section3}, we will identify these operators within the algebra generated by the dual of the central charge operator $\II$. In the next section, we focus on clarifying the structure of the grand-canonical theory. We define it as the infinite direct sum of symmetric orbifolds of all degrees, study the consequences of this definition, and identify the physical basis of states using the language of partial permutations.

\section{The Grand-Canonical Symmetric Orbifold Theory}
\label{Section2}
\subsection{The Hilbert Space}
To construct the Fock space of quantum field theory, one must sum over all $n$-particle Hilbert spaces. Similarly, the Hilbert space of the grand-canonical symmetric orbifold is obtained by summing over symmetric orbifold Hilbert spaces $\curlyH^{(d)}$ of all degrees:
\begin{equation}
    \curlyH = \bigoplus_{d\geq 0} \curlyH^{(d)}\, .
\end{equation}
Each fixed-degree Hilbert space is itself the sum of twisted sectors, which are generated by the twist operators $\sigma_{[g]}^{(d)}$. In the grand-canonical context, we are interested in twists that act at each degree simultaneously, such as
\begin{equation}
    \alpha_k := \bigoplus_{d\geq k} \sigma_{k}^{(d)}
    \label{naivetwist}
\end{equation}
However, there are far too few operators of this form to constitute a basis of the full Hilbert space. To remedy this problem, we generalize the definition \eqref{naivetwist} to an arbitrary partial permutation:
\begin{equation}
    \alpha_{(\curlyS,\rho)} := \bigoplus_{d\geq \lvert \curlyS \rvert} \sigma_{\rho}^{(d)}\, ,
\end{equation}
The corresponding gauge-invariant twists then read
\begin{equation}
    \alpha_r := \bigoplus_{d\geq \lvert d_r \rvert} \binom{m_{1}(\rho)}{m_{1}(r)} \sigma_{\rho}^{(d)}\, .
    \label{mapping}
\end{equation}
where $m_{1}(r)$ is the number of trivial elements in the support of the partial permutation $r$, and $m_{1}(\rho)$ the number of trivial elements in the corresponding permutation $\rho$ of $\{1,2,\dots,d\}$. This combinatorial factor corresponds to the number of partial permutations $(\curlyS,\rho)$ that lift to $\rho$ in $S_d$ under a support-forgetting morphism \cite{IvanovKerov}. The Hilbert space decomposes as
\begin{equation}
    \curlyH = \bigoplus_{[r]} \curlyH_{[r]}\, ,
    \label{Hilbert}
\end{equation}
where $\curlyH_{[r]}$ is the sector of the Hilbert space whose ground state is generated by the twist $\alpha_{r}$, and the sum runs over orbits of partial permutations of all sizes. In this picture, each seed theory operator $X_I$ should be associated with a twist $\alpha_{(I)}$, which means we consider a copy of the seed operator at each degree $d>I$. Furthermore, the identity operator is the partial permutation orbit of empty support $\alpha_{\emptyset}$. 

\subsection{Correlation Functions}
\label{correlators}
We \textit{define} the grand-canonical ensemble of symmetric orbifolds as the direct sum of symmetric orbifolds of all degrees. The one-point function is given by a weighted sum of the one-point function in each theory, weighted by a chemical potential $p$:
\begin{equation}
    \langle \cdot \rangle_p := \sum_d p^d \langle \cdot \rangle^{\rm un}_{d}\, ,
\end{equation}
The recipe for computing correlation functions in the physical basis of operators follows immediately from the definition of partial permutations in terms of fixed-degree twists:
\begin{equation}
    \langle \alpha_{[r]}(z) \cdot (\dots) \rangle_d = \binom{m_{1}(\rho)}{m_{1}(r)} \langle \sigma_{[\rho]}(z) \cdot (\dots) \rangle_d \, .
    \label{combinatorics}
\end{equation}
Geometrically, the operator $\alpha_{[r]}$ selects $\lvert \curlyS_r \rvert$ sheets out of the $d$ available in the covering space. Only those sheets contribute to connected correlators. This geometric picture is perhaps best understood from the dual string perspective, where trivial elements in the support of a partial permutation correspond to disjoint worldsheets that cover the target space exactly once, see section \ref{ConnectedCorrelators}. 

We note that combinatorial differences in fixed-degree correlation functions can lead to drastically different behaviours once summed over all degrees. For example, the operator $\alpha_1$ behaves within any fixed-degree correlation function as a factor of the degree:
\begin{equation}
    \langle \alpha_1 \cdots \rangle_d = d \cdot \langle \cdots \rangle_d \, .
\end{equation}
Nevertheless, its expectation value is non-trivial:
\begin{equation}
    \langle \alpha_1 \rangle = pZ_{seed} \, .
\end{equation}

\subsection{Operator Product Expansion}
The definition and consistency of an associative operator product expansion in the grand-canonical theory follow immediately from its identification as a direct sum of conformal field theories. However, this alone does not guarantee that the coefficients appearing in the expansion of the product of two twist operators $\alpha_r$ are finite, since the change of basis involves infinite sums of states. Their finiteness instead follows from the finiteness of all grand-canonical correlation functions on the sphere. The latter is guaranteed by the fact that the chemical potential decreases exponentially with the degree $d$, whereas symmetric orbifolds correlators are only polynomial in $d$. We stress that these OPE coefficients are independent of the degree, since they were not defined with respect to any specific value of $d$. 

Although the existence of the operator product expansion is guaranteed, it is only natural to ask for an intrinsic definition in the language of partial permutations. We now provide such a definition. Recall that to each seed theory field $X_I$, we associate a twist operator $\alpha_{(I)}$. Therefore, given the OPE coefficients of the seed theory, the problem reduces to computing products of twist operators $\alpha_{(\curlyS_1,\rho_1)}$ and $\alpha_{(\curlyS_2,\rho_2)}$, and regrouping the resulting terms. The product lies in a single channel $\alpha_{(\curlyS_1\cup \curlyS_2,\sigma_{\rho_1} \sigma_{\rho_2})}$. Within this channel, the various contributions are obtained by embedding the twist operators into the infinite symmetric orbifold and computing the operator product expansion of $\sigma_{\rho_1}$ with $\sigma_{\rho_2}$. In fact, the computation may be carried out in a symmetric orbifold of any degree $d\geq \lvert \curlyS_1 \rvert + \lvert \curlyS_2 \rvert - 1$. Indeed, its result is independent of spectator copies, that lie outside the support of the twists. Finally, one groups together appropriate contributions from each channel.

A simple algorithm consists of first computing the operator product expansion at some fixed degree, then replacing each factor of the degree by an operator $\alpha_1$, and finally multiplying out the factors of $\alpha_1$ using equations \eqref{algebra} and \eqref{singlecyclealgebra}. Equivalently, polynomials of the degree may be decomposed in the basis of binomial polynomials $p_k(d):=\binom{d}{k}$, and each $p_k$ replaced with an operator $\alpha_{1^k}$. For instance, given a free scalar field as seed theory, the operator product expansion of the field $X:=\sum_I X_I$ with itself reads
\begin{align*}
    \pd X (z)\, \pd X (w) = &-\frac{d}{(z-w)^2}+ \dots\\
    \rightarrow &-\frac{\alpha_1}{(z-w)^2}+ \dots\, ,
\end{align*}
seeing as $\alpha_1$ behaves as a factor of the degree in any fixed-degree correlation function. In our framework, one associates a trivial twist $\alpha_{(I)}$ to each copy of the seed field $X_I$, and the resulting $\alpha_1$ arises from the sum over all $\alpha_{(I)}$. To conclude this section, we note that the partial permutation framework offers an illuminating perspective on operator product expansions in the large $N$ limit of symmetric orbifolds \cite{Ashok:2023kkd}. 

\subsection{Virasoro Algebra}
The stress-energy tensor of the grand-canonical theory reads
\begin{equation}
    T_{\scaleto{GC}{4pt}}(z)=\sum_{I\geq 0} T_{I}(z) \alpha_{I}\, ,
    \label{Hamiltonian}
\end{equation}
where $T_{I}$ denotes the stress-energy tensor of the $I$-th copy. Indeed, this is a conserved spin-$2$ tensor that generates conformal transformations, because the truncation to degree $d$ of $T(z)$ is the fixed-degree stress tensor $T_d$. We write
\begin{align*}
    \langle \left[\frac{1}{2\pi i} \int dz \, T(z) \epsilon(z),\Phi(w)\right] \cdots \rangle &= \sum_d p^d \langle \delta_{\epsilon}\Phi(w) \cdots \rangle_d\\
    &=\langle \delta_{\epsilon}\Phi(w) \cdots \rangle\, .
\end{align*}
The operator product expansion of the stress tensor with itself reads
\begin{equation}
    T(z) T(0) = \frac{(c_{seed}/2)\cdot \alpha_1}{(z-w)^4}+\frac{T(z)}{(z-w)^2}+\frac{\pd T(z)}{z-w}+\text{regular}\, .
\end{equation}
The central charge is represented non-trivially as the operator $c_{seed}\, \alpha_1$. The corresponding representation of the Virasoro algebra reads
\begin{equation}
    [L_n,L_m]=(n-m)L_{n+m}+\frac{c_{seed}\, \alpha_1}{12}(n^3-n) \delta_{n+m,0}\, ,
    \label{Virasoro}
\end{equation}
where the $L_n$ are the Laurent modes of the stress-energy tensor, $L_n:=\int dz\, z^{n+1} T(z)$. An infinite-dimensional generalization of Schur's lemma implies that the representation at hand is reducible, as it admits a non-scalar central element. This is consistent with the definition of the grand-canonical ensemble as a direct sum of conformal field theories with distinct central charges. 

The appearance of a non-scalar central charge in the representation \eqref{Virasoro} is consistent with the findings of \cite{Troost:2010zz} for the spacetime Virasoro algebra of $AdS_3$. In the tensionless context, it leads to the identification of $\alpha_1$ as the counterpart of the central charge operator $\II$ of the dual string theory. The expectation value of $c_{seed}\, \alpha_1$ therefore computes the Brown-Henneaux central charge \cite{Troost:2011ud}. This allows us to identify the relation between the string coupling and the chemical potential. The three-dimensional Newton constant for an $AdS_3\times S^3 \times T^4$ background at $k=1$ reads \cite{Polchinski:1998rr} \linebreak[4]
\vspace{-\baselineskip}
\begin{equation*}
    G_N^{(3)}=\frac{G_N^{(10)}}{\text{Vol}(S^3) \cdot \text{Vol}(\mathbb{T}^4)} = \frac{8 \pi^6 g_s^2 R^8}{2\pi^2 R^3\cdot (2\pi)^4 v_4 R^4}\, ,
\end{equation*}
where $v_4=\frac{\text{Vol}(\mathbb{T}^4)}{(2\pi R)^4}$. The Brown-Henneaux central charge is thus
\begin{equation}
    c_{\scaleto{BH}{4pt}}=\frac{3\ell}{2G_N^{(3)}}=\frac{6 v_4}{g_{s}^{2}}\, ,
\end{equation}
which leads to
\begin{equation}
    pZ_{seed}=v_4\cdot g_{s}^{-2}\, ,
\end{equation}
consistent with \cite{Aharony:2024fid}. We note however that the partition function of the seed conformal field theory has an ambiguity \cite{Gerchkovitz:2014gta,Eberhardt:2023lwd}. 

At this stage, let us summarize our construction. We have shown that the grand-canonical ensemble of symmetric orbifolds can be defined as the direct sum of orbifolds of all degrees. This ensures that the resulting theory is an ordinary quantum field theory, with a well-defined and associative operator product expansion. It is also unitary, provided the seed theory is. Crucially, the basis of operators that admits a physical interpretation mixes all sectors. We have explained how to take operator product expansions in this basis, except for two formulas describing the combinatorics of the operator $\alpha_1$. We establish them in the next section.

\section{Central Operators}
\label{Section3}
The generalized twist operator $\alpha_1$ is of special interest. Indeed, it represents the central charge of the Virasoro algebra of the grand-canonical theory. Furthermore, together with the standard twist operators $\alpha_k$ and the seed excitations, it generates the entire spectrum of the theory. In this section, we study some of its properties.

\subsection{The Central Charge Operator}
The operator $\alpha_1(x)$ is independent of its position $x$. This follows immediately from equation \eqref{combinatorics}. We will therefore consistently omit its insertion point. Furthermore, $\alpha_1$ acts on unnormalized correlators as the differential operator $\pd_{\log p}$:
\begin{align}
    \langle \alpha_1 \cdot (\dots) \rangle^{\rm un} & := \sum_{d\geq 1} d\, p^d \langle (\dots) \rangle^{\rm un} _d \\
    &= p\pd_{p} \langle (\dots) \rangle^{\rm un}\, .
    \label{CentralAction}
\end{align}
This behaviour is consistent with that of the central charge operator $\II$ of an $AdS_3$ string \cite{Eberhardt:2023lwd,Aharony:2024fid}. In particular, it implies that inserting the latter operator in a disconnected, fixed-degree correlator multiplies it by a factor of the degree.

\subsection{The Universal Subalgebra}
\label{Subalgebra}
Let us now study the tower of operators $\alpha_{1^k}$, all of which are once again independent of their insertion point. Moreover, they are central, meaning that they commute with every operator in the theory. Their vacuum expectation value is 
\begin{align}
    \langle \alpha_{1^k} \rangle 
    &=\frac{p^k Z_{seed}^k}{k!}\, .
\end{align}
Furthermore, their action within any correlation function reads
\begin{align}
    \langle \alpha_{1^k} \cdot (\dots) \rangle^{\rm un} & := \sum_{d\geq 1} \binom{d}{k} p^d \langle (\dots) \rangle^{\rm un} _d\, , \\
    &= \frac{p^k \pd_p^k}{k!} \langle (\dots) \rangle^{\rm un} \, .
\end{align}
Therefore, we identify
\begin{equation}
    \alpha_{1^k}\equiv \frac{p^k \pd_p^k}{k!}\, .
\end{equation}
The algebra of central operators is isomorphic to the algebra of differential operators $p^k \pd_p^k/k!$. It can be obtained by induction. To begin with, we compute
\begin{equation}
    \alpha_{1^{k+1}}\alpha_{1^l} = \frac{1}{k+1}\,  (\alpha_1-k) \cdot \alpha_{1^k} \alpha_{1^l}\, .
    \label{semirecursion}
\end{equation}
We denote by $a_{kl}^{\ \ m}$ the coefficients of $\alpha_{1^m}$ in the product $\alpha_{1^k}\cdot \alpha_{1^l}$. Equation \eqref{semirecursion} leads to the following recursion relation for the coefficients $a_{kl}^{\ \ m}$:
\begin{equation}
    a_{k+1,l}^{\quad \quad n} = \frac{1}{k+1}\left[(n-k) a_{kl}^{\ \ n} + n a_{kl}^{\ \ n-1} \right]\, ,
\end{equation}
By solving this relation, we obtain the following algebra of central operators
\begin{equation}
    \alpha_{1^k}\alpha_{1^l} = \sum_{m=0}^{\min(k,l)} \frac{(k+l-m)!}{m! (k-m)! (l-m)!} \; \alpha_{1^{k+l-m}}\, .
    \label{algebra}
\end{equation}
By construction, this algebra is isomorphic to the freely generated algebra of the single element $\II$. The content of equation \eqref{algebra} is combinatorial: in the basis most appropriate to our problem, the structure constants take the form above. Furthermore, the fact that operators generated by $\II$ define an algebra is consistent with the independence of this operator on its position in spacetime. 

Let us also point out that this algebra is isomorphic to an algebra of creation and annihilation operators whose basis is given by the $W_k:=\frac{1}{k!}(a^{\dagger})^k a^k$, as shown in \cite{Blasiak}. This suggests that an intuitive way to think about the operator $\II$ is as a number operator $N=a^{\dagger} a$. Furthermore, as shown in \cite{Blasiak}, taking powers of the central charge operator $\II$ computes Stirling numbers of the second kind $\genfrac\{\}{0pt}{}{n}{m}:=\frac{1}{m!}\sum_{j=0}^{m}\binom{m}{j}(-1)^{m-j} j^n$:
\begin{equation}
    (\alpha_1)^n = \sum_m m!\, \genfrac\{\}{0pt}{}{n}{m} \cdot  \alpha_{1^m}\, .
    \label{Powers}
\end{equation}
The formula \eqref{Powers} effectively identifies the polynomial $P_{n}(x)$ that appeared in \cite{Kim:2015gak}.
       
\subsection{OPE with the Central Charge Operator}
\label{CentralOPE}
The action of the central charge operator within any fixed-degree correlator is purely  combinatorial. Therefore, the operator product expansion of any power of the central charge with a generalized twist reduces to their product as orbits of partial permutations \cite{IvanovKerov}. To compute it, we first write
\begin{equation}
    \alpha_k \cdot \alpha_{1^{l}} = \sum_{i=0}^{\min(k,l)} \binom{k}{i} \alpha_{k,1^{l-i}}\, .
\end{equation}
We then expand $\alpha_1^n$ using equation (31), plug the above expression in the product and regroup terms. For simplicity, we will assume that $k\geq n$. We find
\begin{equation}
    \alpha_k \cdot \alpha_{1}^n =  \sum_{p=0}^{n} \left[\sum_{l=p}^{n}  l!  \genfrac\{\}{0pt}{}{n}{l} \binom{k}{l-p}\right] \alpha_{k,1^{p}}\, .
    \label{singlecyclealgebra}
\end{equation}
Stirling numbers of the second kind are generated by the exponential function  $\frac{1}{l!}(e^x-1)^l$. Therefore, the coefficient $a^{(k,n)}_{p}$ of $\alpha_{k,1^p}$ is given by
\begin{align}
    a^{(k,n)}_{p} &=n! [x^n] e^{kx}(e^x-1)^p\\
    &=\sum_{j=0}^{p} (-1)^{p-j} \binom{p}{j} (k+j)^n \, .
\end{align}
The simplest example is the operator product expansion of the central charge operator with a single cycle twist. It reads
\begin{equation}
    \alpha_1 \alpha_k=\alpha_{k,1}+k \alpha_k\, ,
    \label{IOPE}
\end{equation}
with $k\geq 2$. A geometric interpretation of equation \eqref{IOPE} is that $\alpha_1$ either marks an existing sheet or creates a new one. Holographically, the latter corresponds to adding a disjoint worldsheet in its vacuum state. A key property of the Hilbert space \eqref{Hilbert} is indeed that it distinguishes between vacuum configurations with different numbers of worldsheets in their vacuum state. The string number is equal to the number of partial permutation orbits in the dual state. For instance, $\alpha_{1,3}=\alpha_{(1)(234)}+\text{etc.}$ represents a two-string state.  Our construction suggests that for a boundary description to exhibit a field theory structure, and an operator product expansion in particular, it must holographically encode a multi-string Hilbert space.

\subsection{Connected Correlation Functions}
\label{ConnectedCorrelators}
Connected correlation functions are obtained by restricting the sum over covering surfaces to connected coverings. The surfaces relevant to this definition are determined solely by the partial permutations that appear in the correlator. Hence, we say that a covering space is connected if the corresponding set of partial permutations acts transitively on its support. For example, $\langle \alpha_{(12)(3)} \alpha_{(23)} \rangle$ gives rise to a connected covering, whereas $\langle \alpha_{(12)(4)} \alpha_{(23)} \rangle$ does not. Consequently, the action of the central charge operator in connected correlators reads
\begin{equation}
    \langle \alpha_1 \cdot \prod_{i} \alpha_{r_i} \rangle_{d}^{\rm conn} = N_c \cdot  \langle \prod_{i} \alpha_{r_i} \rangle_{d}^{\rm conn}\, ,
\end{equation}
where $N_c$ is the number of connected sheets in the surface defined by the partial permutations $r_i$. Indeed, as previously explained, the operator $\alpha_1$ either marks a sheet or adds a new one, and only the former action contributes in the connected case. In fact, this behaviour follows directly from our identification of $\alpha_1$ as the dual of the central charge operator $\II$, as shown in \cite{Kutasov:1999xu}. It is also consistent with our interpretation of $\alpha_1$ as the analogue of a particle number. 

Finally, we note that on the sphere, disconnected single-sheeted contributions can be factored out of any worldsheet and cancelled against the full partition function. This is akin to the argument used to discard bubble diagrams in quantum field theory. It implies that connected grand-canonical correlators on the sphere reduce to
\begin{equation}
    \langle \prod_{i} \alpha_{k_i} \rangle^{\rm conn}_p = \sum_{N_c} p^{N_c} \cdot \langle \prod_{i} \alpha_{k_i} \rangle^{\rm un,\,  conn}_{N_c}\, ,
\end{equation}
where the sum runs over values $N_c$ of the degree for which connected covers without extra disconnected spheres exist. The set of allowed values is finite, as it is constrained by the Riemann-Hurwitz formula.
This offers yet another perspective on the advantage of going grand-canonical: it eliminates the need to account for disconnected sphere diagrams.

\section{Outlook}
In this letter, we have shown that grand-canonical ensembles of symmetric orbifold conformal field theories can be formulated as the direct sum of symmetric orbifolds of all degrees. As such, they define unitary and conformally invariant quantum field theories that admit a consistent operator product expansion. Our formulation is best suited to describe a second-quantized Hilbert space for the dual string theory. In this context, we have provided a conformal field theory interpretation of the central charge operator of an $AdS_3$ string theory at the tensionless point. It would be interesting to understand the field-theoretic origin of the central charge operator beyond the unit-flux case. String theory on an $AdS_3\times X$ background with pure NS-NS flux is believed to be dual to a deformed symmetric orbifold in the grand-canonical ensemble \cite{Eberhardt:2021vsx,Knighton:2024pqh}. We therefore expect the ideas developed here to be of relevance in the more general case. 

Another avenue for future work would be to generalize our construction to symmetric orbifolds with boundaries. This may shed light on some grand-canonical aspects of holographic interfaces in $AdS_3/CFT_2$ \cite{Gaberdiel:2021kkp,Knighton:2024noc,Harris:2025wak,Belleri:2025eun,Harris:2025klp}. A first step in this direction was taken in \cite{Troost:2025eqm}, which explores grand-canonical Hurwitz theories with boundaries.\linebreak[4]

\vspace{-7mm}

\acknowledgments{It is a pleasure to thank Jan Troost for many stimulating discussions and for comments on an early version of the manuscript. I am also grateful to Ant\'onio  Antunes, Simon Douaud, Sebastian Harris and Eric Perlmutter for helpful conversations. I thank the anonymous referee for comments that led to improvements and corrections, and DESY for its hospitality during the period in which this work was revised.} 

\bibliographystyle{unsrt}
\bibliography{bib}

\end{document}